\documentclass[aps,pre,
preprint,
tightenlines,
showpacs]{revtex4}

\usepackage{graphicx}

\begin{document}

\title{
Avoided crossings in three coupled oscillators as a model system of acoustic 
bubbles
}

\author{Masato Ida}
\email[E-Mail: ]{ida.masato@jaea.go.jp}
\thanks{Present address: Center of Computational Science and Engineering, 
Japan Atomic Energy Agency, 6-9-3 Higashi-Ueno, Taito-ku, Tokyo 110-0015, Japan}
\affiliation{
Center for Promotion of Computational Science and Engineering, Japan Atomic 
Energy Research Institute,
6-9-3 Higashi-Ueno, Taito-ku, Tokyo 110-0015, Japan
}

\begin{abstract}
The resonance frequencies and oscillation phases of three acoustically 
coupled bubbles are examined to show that avoided crossings can appear in a 
multibubble system. Via a simple coupled oscillator model, we show that if 
at least three bubbles exist, it is possible for their resonance frequencies 
as functions of the separation distances between the bubbles to experience 
an avoided crossing. Furthermore, by focusing our attention on the 
oscillation phases and based on analysis of the transition frequencies [M.~Ida, 
Phys.~Lett.~A {\bf 297}, 210 (2002); M.~Ida, J.~Phys.~Soc.~Jpn.~{\bf 71}, 1214 
(2002)] of the coupled bubbles, we show that a distinct state exchange takes 
place between the bubbles at a point in the avoided crossing region, where a 
resonance frequency of the triple-bubble system crosses with a transition 
frequency not corresponding to the resonance frequencies.
\end{abstract}

\pacs{47.55.Dz; 43.20.+g; 47.55.Bx}

\maketitle

\section{Introduction}
\label{sec1}
Avoided crossings \cite{ref1} have been observed theoretically and experimentally in 
a large variety of physical systems involving eigenvalues (e.g., natural 
frequencies, eigenenergies) \cite{ref2,ref3,ref4,ref5,ref6,ref7,ref8,ref9,ref10,ref11,ref12,ref13,ref14,ref15}, and they have attracted much attention even 
in recent years because of their rich physics and practical importance in, 
for example, mechanical engineering \cite{ref11,ref12,ref13,ref14} and quantum physics \cite{ref2,ref3,ref4,ref5}. In the 
avoided crossing regions, eigenvalues of the system first approach each 
other as a system parameter is varied but then veer abruptly from each other 
without crossing. In those regions a drastic change of some characteristic 
of the system occurs along the eigenvalue loci. In Ref.~\cite{ref12}, for 
example, Pierre illustrated that the mode shapes of a disordered chain of 
coupled pendulums change in the regions where avoided crossings of the 
eigenfrequencies of the system take place. In that study, disorders in the 
lengths of the pendulums were used as the system parameters. Also, in 
Ref.~\cite{ref2}, Walkup et al.~studied in detail avoided crossings observed in the 
energy levels of diamagnetic hydrogen as functions of the magnetic field 
strength or the angular momentum, which lead to the diabatic exchange of the 
states of the wave functions. A study of Bose-Einstein condensation (BEC) 
\cite{ref5} showed that in order to trap molecules created in an atomic BEC through 
a Feshbach resonance, an avoided crossing of two bound states of the 
molecules must be exploited, through which the vibrational quantum number 
and size of the trapped molecules change.

In the present paper, we show theoretically that avoided crossings can be 
observed in acoustically coupled bubbles, which has to the authors' 
knowledge not been stated in the literature. Furthermore, based on analyses 
of transition frequencies \cite{ref16,ref17}, we propose a way to detect a state 
exchange occurring in the avoided crossing region. The theoretical model 
used in this study, reviewed in Sec.~\ref{sec2}, is a forced coupled oscillator 
model that describes acoustic coupling of pulsating bubbles. Using the 
model, we show in Sec.~\ref{sec3} that if at least three bubbles exist, it is 
possible that the resonance frequencies of the bubbles exhibit an avoided 
crossing when they are plotted as functions of the separation distances 
between the bubbles. As has been demonstrated (e.g., Refs.~\cite{ref16,ref18}), in 
double-bubble systems, neither crossings nor avoided crossings of the 
resonance frequencies as functions of the separation distance are observed, 
since the higher of the two resonance frequencies of the systems increases 
and the lower one decreases as the bubbles approach each other. However, as 
shown in the present paper, by introducing one more bubble whose monopole 
(i.e., decoupled) resonance frequency crosses with one of the resonance 
frequencies of a double-bubble system, one can observe the avoided crossing 
of the resonance frequencies when all three bubbles are coupled.

In Sec.~\ref{sec4}, we examine the phase properties of the three coupled bubbles to 
show that a state exchange actually occurred between the bubbles in the 
avoided crossing region. In this effort, the notion of a transition 
frequency plays an important role. The transition frequencies introduced in 
Refs.~\cite{ref16,ref17} are characteristic frequencies of acoustically coupled 
bubbles, around which the oscillation phase of bubbles inverts, e.g., from 
in-phase to out-of-phase with the driving sound. It was proved in 
Ref.~\cite{ref17} that a bubble in a $N$-bubble system has up to $2N-1$ transition 
frequencies, only $N$ ones of which correspond to the resonance frequencies 
of the system. That is, observing the transition frequencies allows us to 
obtain richer insight into the phase properties than that obtained by only 
observing the resonance frequencies. This notion has already been exploited 
as a powerful tool to understand the sign reversal of the secondary Bjerknes 
force \cite{ref18,ref19} in which the oscillation phases play a crucial role. Using 
this notion and observing directly the oscillation phases, we show that the 
coupled bubbles exchange their oscillation states through the avoided 
crossing and the state exchange takes place at the separation distances 
where an avoided crossing resonance frequency crosses with a transition 
frequency that is not a resonance frequency. The present findings appear to 
reveal a taste of bubbles' hidden complexity.

Section \ref{sec5} summarizes this paper, and the Appendixes present additional 
remarks.

\section{Coupled oscillator model, resonance frequency, and transition 
frequency}
\label{sec2}
The theoretical model used in the present study is a forced oscillator model 
in which $N$ harmonic oscillators are coupled (\cite{ref16,ref17} and 
references therein):
\begin{eqnarray}
\label{eq1}
\ddot {e}_i +\omega _{i0}^2 e_i +\delta _i \dot {e}_i &=& -\frac{p_{{\rm ex}} 
}{\rho R_{i0} }-\frac{1}{R_{i0} }\sum\limits_{j=1,\;j\ne i}^N 
{\frac{R_{j0}^2 }{D_{i\,j} }\ddot {e}_j } \\ 
\mbox{for}\;\;i &=& 1,2,\cdots ,N, \nonumber
\end{eqnarray}
where $N$ corresponds to the number of bubbles, $R_{i0} $ is the equilibrium 
radius of bubble $i$, $e_i $ is the deviation of radius assumed as $\left| 
{e_i } \right|\ll R_{i0} $, $\omega _{i0} $ is the monopole (angular) 
resonance frequency of bubble $i$, defined as
\begin{equation}
\label{eq2}
\omega _{i0} =\sqrt {\frac{3\kappa _i P_0 +(3\kappa _i -1)2\sigma /R_{i0} 
}{\rho R_{i0}^2 }} ,
\end{equation}
$\delta _i $ is the damping factor, the overdots denote the time derivation, 
$p_{{\rm ex}} $ is the pressure of the external sound, $\rho $ is the 
density of the surrounding liquid, $D_{i\,j} $ ($=D_{j\,i} )$ is the 
separation distance between the centers of bubbles $i$ and $j$, $\kappa _i $ 
is the polytropic exponent of the gas inside the bubbles, $P_0 $ is the 
static pressure, and $\sigma $ is the surface tension. In this linear model, 
the following assumptions are made: the surrounding liquid is 
incompressible, the sound amplitude is sufficiently low, the separation 
distances are much larger than the bubbles' radii, and the shape deformation 
of the bubbles is negligible. The last term of Eq.~(\ref{eq1}), representing the 
pressures of the sounds that the neighboring bubbles emit, describes the 
acoustic coupling between the bubbles. As in the double-bubble case 
\cite{ref20}, this model may be assumed to be of third order with respect to 
the inverse of the separation distances (i.e., the truncated terms are of 
fourth or higher order); see Appendix \ref{secA}.

Using this model with $N=3$, a matrix equation for determining the 
amplitudes and phases of the radial oscillations is derived. Assuming 
$p_{{\rm ex}} =-P_a \exp ({\rm i}\omega t)$ and $e_i =\beta _i \exp 
({\rm i}\omega t)$ with $P_a $ being a positive constant, $\omega $ being 
the driving (angular) frequency, and $\beta _i $ being a complex amplitude, 
we have
\begin{equation}
\label{eq3}
{\rm {\bf A}}\left( {{\begin{array}{*{20}c}
 {\beta _1 } \hfill \\
 {\beta _2 } \hfill \\
 {\beta _3 } \hfill \\
\end{array} }} \right)=-\frac{P_a }{\rho }{\rm {\bf I}},
\end{equation}
where ${\rm {\bf A}}$ is a $3\times 3$ matrix whose elements, $a_{i,j} $ 
($i,j=1,2,3)$, are defined as
\begin{equation}
\label{eq4}
a_{i,j} \equiv \left\{ {{\begin{array}{*{20}c}
 {R_{i0} [(X-\omega _{i0}^2 )-{\rm i}\omega \delta _i ]} & 
{\mbox{for}\;\;i=j,} \\
 {\displaystyle{\frac{R_{j0}^2 }{D_{i\,j} }X}} & {\mbox{otherwise,}} \\
\end{array} }} \right.
\end{equation}
with
\begin{equation}
\label{eq5}
X\equiv \omega ^2,
\end{equation}
and ${\rm {\bf I}}=(1,1,1)^T$. We should note here that essentially the same 
matrix equations can be found in previous papers (e.g.~\cite{ref21,ref22,ref17}). The solution of Eq.~(\ref{eq3}) is represented as
\begin{eqnarray}
\label{eq6}
\left( {{\begin{array}{*{20}c}
 {\beta _1 } \hfill \\
 {\beta _2 } \hfill \\
 {\beta _3 } \hfill \\
\end{array} }} \right) &=& -\frac{P_a }{\rho }{\rm {\bf A}}^{-1}{\rm {\bf I}} \\ \nonumber
&=& -\frac{P_a }{\rho }\frac{\vert {\rm {\bf A}}\vert ^\ast {\rm {\bf 
C}}\,{\rm {\bf I}}}{\vert {\rm {\bf A}}\vert ^\ast \vert {\rm {\bf A}}\vert },
\end{eqnarray}
where $\vert {\rm {\bf A}}\vert $ and ${\rm {\bf C}}$ are the determinant 
and the cofactor matrix of ${\rm {\bf A}}$, respectively, and $\vert {\rm 
{\bf A}}\vert ^\ast $ is the complex conjugate of $\vert {\rm {\bf A}}\vert 
$. We used here an expression in which the denominator is real.

The eigenfrequencies of the system are determined by
\begin{equation}
\label{eq7}
\vert {\rm {\bf A}}\vert =0,
\end{equation}
which is a cubic equation in terms of $X$. For $\delta _i \approx 0$, the 
roots of this equation are equivalent to the resonance frequencies of the 
system. The transition frequencies of bubble $i$, defined as the driving 
frequencies at which the phase difference between bubble $i$ and the driving 
sound is $\pi /2$ (or $3\pi /2)$ \cite{ref16,ref17,ref18}, are determined by
\begin{equation}
\label{eq8}
{\rm Re}(\tau _i )=0,
\end{equation}
where
\begin{equation}
\label{eq9}
\left( {{\begin{array}{*{20}c}
 {\tau _1 } \hfill \\
 {\tau _2 } \hfill \\
 {\tau _3 } \hfill \\
\end{array} }} \right)\equiv \vert {\rm {\bf A}}\vert ^\ast {\rm {\bf 
C}}\,{\rm {\bf I}}.
\end{equation}
(See Appendix \ref{secB} for the concrete forms of $\left| {\rm {\bf A}} \right|$ and 
${\rm {\bf C}}\,{\rm {\bf I}}$.) From the mathematical proof given in 
Ref.~\cite{ref17}, one knows that Eq.~(\ref{eq8}) is a fifth-order polynomial 
in terms of 
$X$, meaning that the bubbles may have up to five transition frequencies.

The phase delay of bubble $i$, denoted by $\phi _i $, measured from the 
phase of the driving sound is determined using the ${\rm atan2}(y,x)$ 
function in the C language, which returns $\tan ^{-1}(y/x)\in [-\pi ,\pi ]$, 
as
\[
\phi _i =\left\{ {{\begin{array}{*{20}l}
 {\psi _i } & {\mbox{if }\psi _i \ge 0,} \\
 {\psi _i +2\pi } & {\mbox{otherwise}} \\
\end{array} }} \right.
\]
with
\[
\psi _i ={\rm atan2}(-{\rm Im}(\tau _i ),{\rm Re}(\tau _i )).
\]
The next section shows that in certain cases an avoided crossing is observed 
in the solution of Eq.~(\ref{eq7}). In the discussion, to obtain real 
eigenfrequencies that correspond to the resonance frequencies of the 
triple-bubble system for weak damping, we for the moment assume $\delta _i 
\approx 0$ (but $\delta _i \ne 0)$. Under this assumption, one obtains
\begin{equation}
\label{eq10}
{\rm Im}(\vert {\rm {\bf A}}\vert )\approx {\rm {\bf 0}},
\end{equation}
\begin{equation}
\label{eq11}
\vert {\rm {\bf A}}\vert \approx \vert {\rm {\bf A}}\vert ^\ast ,
\end{equation}
and
\begin{equation}
\label{eq12}
\tau _i \approx {\rm Re}(\tau _i ).
\end{equation}
Influences of the damping effect on the phase properties are briefly 
discussed in Sec.~\ref{sec4}.

\section{Avoided crossings of resonance frequencies}
\label{sec3}
\begin{figure}[htbp]
\includegraphics[width=7cm]{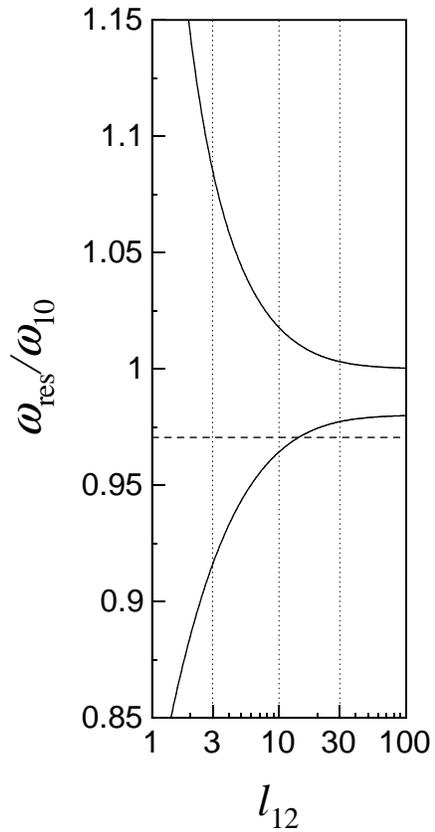}
\caption
{Resonance frequencies $\omega _{{\rm res}} $ (rad/s) of two coupled 
bubbles for $\delta _i \approx 0$ normalized by $\omega _{10} $ (rad/s), as 
functions of the normalized separation distance $l_{12} $. The dashed line 
denotes the monopole resonance frequency of a bubble that will be coupled 
with the former two bubbles in the next example.}
\label{fig1}
\end{figure}

To begin with, a double-bubble system is briefly reconsidered to confirm 
that no avoided crossings are observed in the resonance frequencies of the 
system as functions of the separation distance. The solid lines in 
Fig.~\ref{fig1} 
indicate the resonance frequencies of two coupled bubbles (bubbles 1 and 2) 
of $(R_{10} ,R_{20} )=(50\;{\rm \mu m},51\;{\rm \mu m})$ as functions of 
$l_{12} =D_{12} /(R_{10} +R_{20} )$. The other parameters are set to 
$\rho =1000$ kg/m$^3$, $\kappa _i =1.4 \quad (i=1,2,3)$, $P_0 =1$ atm, and 
$\sigma =0.0728$ N/m. As has been proved theoretically \cite{ref23,ref24,ref16,ref20}, two resonance (or natural) frequencies appear in this system, 
each of which, for $D_{12} \to \infty $, converges to the monopole 
resonance frequency of a bubble. The higher resonance frequency increases 
and the lower decreases as the separation distance decreases. It is 
therefore obvious that avoided crossings cannot occur.

Here we introduce one more bubble into the system. The dashed line displayed 
in Fig.~\ref{fig1} denotes the monopole resonance frequency of the introduced bubble, 
bubble 3, whose radius $R_{30} =51.5\;{\rm \mu m}$. Note that this 
resonance frequency crosses with a resonance frequency of the double-bubble 
system. This crossing, as shown immediately, triggers an avoided crossing 
when the third bubble is coupled with the double-bubble system.

\begin{figure}[htbp]
\includegraphics[width=14cm]{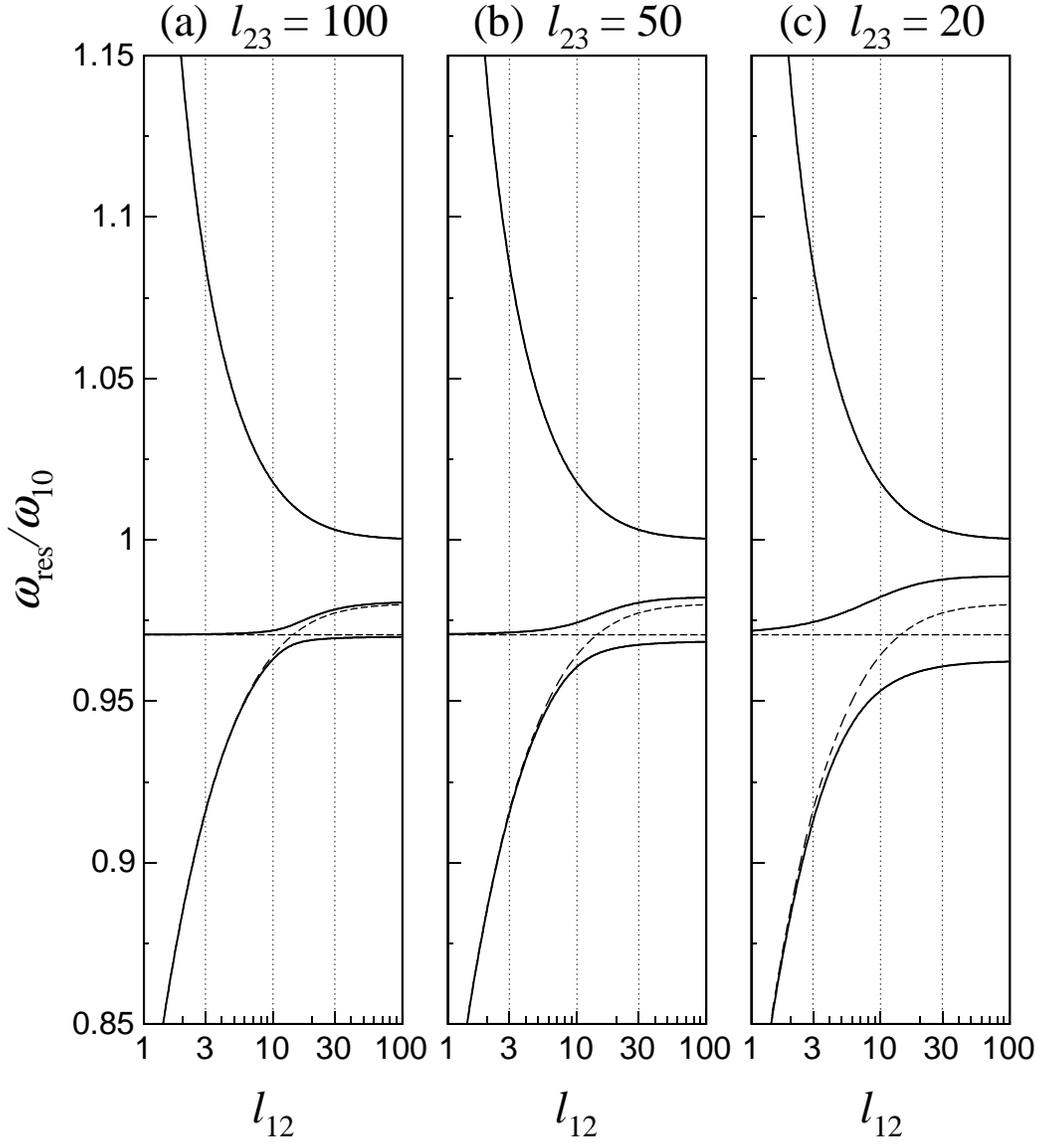}
\caption
{Resonance frequencies $\omega _{{\rm res}} $ (rad/s) of three 
coupled bubbles for $\delta _i \approx 0$ normalized by $\omega _{10} $ 
(rad/s), as functions of the normalized separation distance $l_{12} $. 
Figures \ref{fig2}(a), \ref{fig2}(b), and \ref{fig2}(c) are for $l_{23} =100$, $50$, and $20$, 
respectively. The dashed lines denote the resonance frequencies when bubble 
3 is decoupled.}
\label{fig2}
\end{figure}

\begin{figure}[htbp]
\includegraphics[width=12cm]{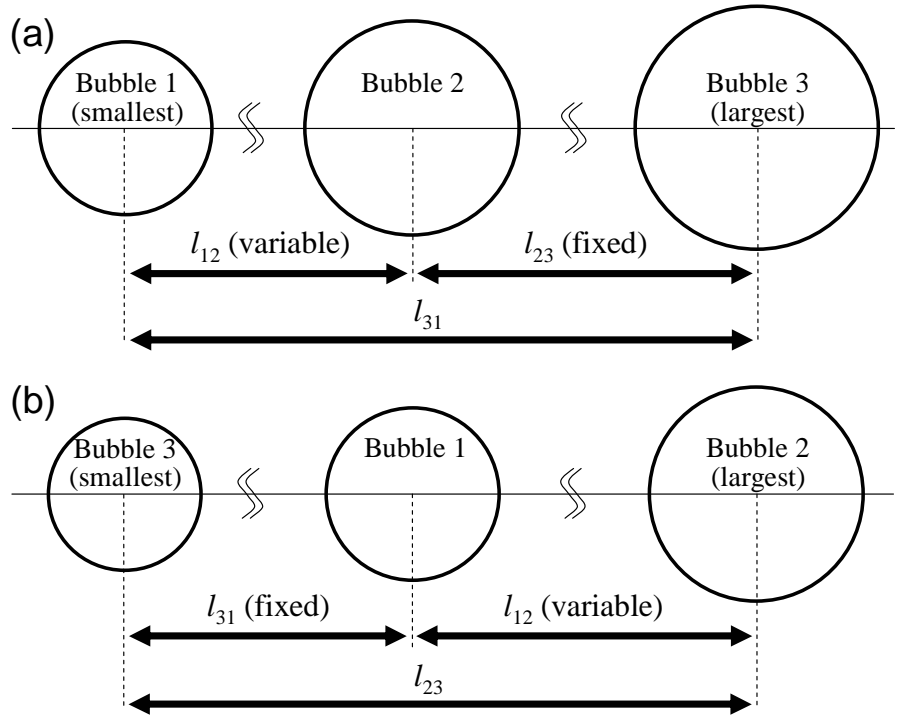}
\caption
{Arrangements of bubbles in the cases where bubble 3 is larger (a) 
and smaller (b) than the other two bubbles.}
\label{fig3}
\end{figure}

Figure \ref{fig2} shows the resonance frequencies in the case where all three bubbles 
are coupled. The separation distances are set to $D_{12} =l_{12} (R_{10} 
+R_{20} )$, $D_{23} =l_{23} (R_{20} +R_{30} )$, and $D_{31} =D_{12} 
+D_{23} $; that is, the bubbles are arranged in line (see Fig.~\ref{fig3}(a)). Here 
the nondimensional quantities $l_{12} $ and $l_{23} $ are used as the system 
parameters. Figures \ref{fig2}(a), \ref{fig2}(b), and \ref{fig2}(c) show the results for $l_{23} =100$ \cite{ref25}, $50$, and $20$, respectively. In the 
figures, an avoided crossing is 
clearly seen that takes place around the point at which the two decoupled 
resonance frequencies cross. The line of the resonance frequency originating 
with bubble 3 is divided into two parts, and each of them connects smoothly, 
like blending, with the curve of a resonance frequency of the double-bubble 
system, also divided into two parts. As bubble 3 comes closer to the others, 
the avoided crossing becomes broader and the origin of each resonance 
frequency becomes increasingly unclear.

\begin{figure}[htbp]
\includegraphics[width=14cm]{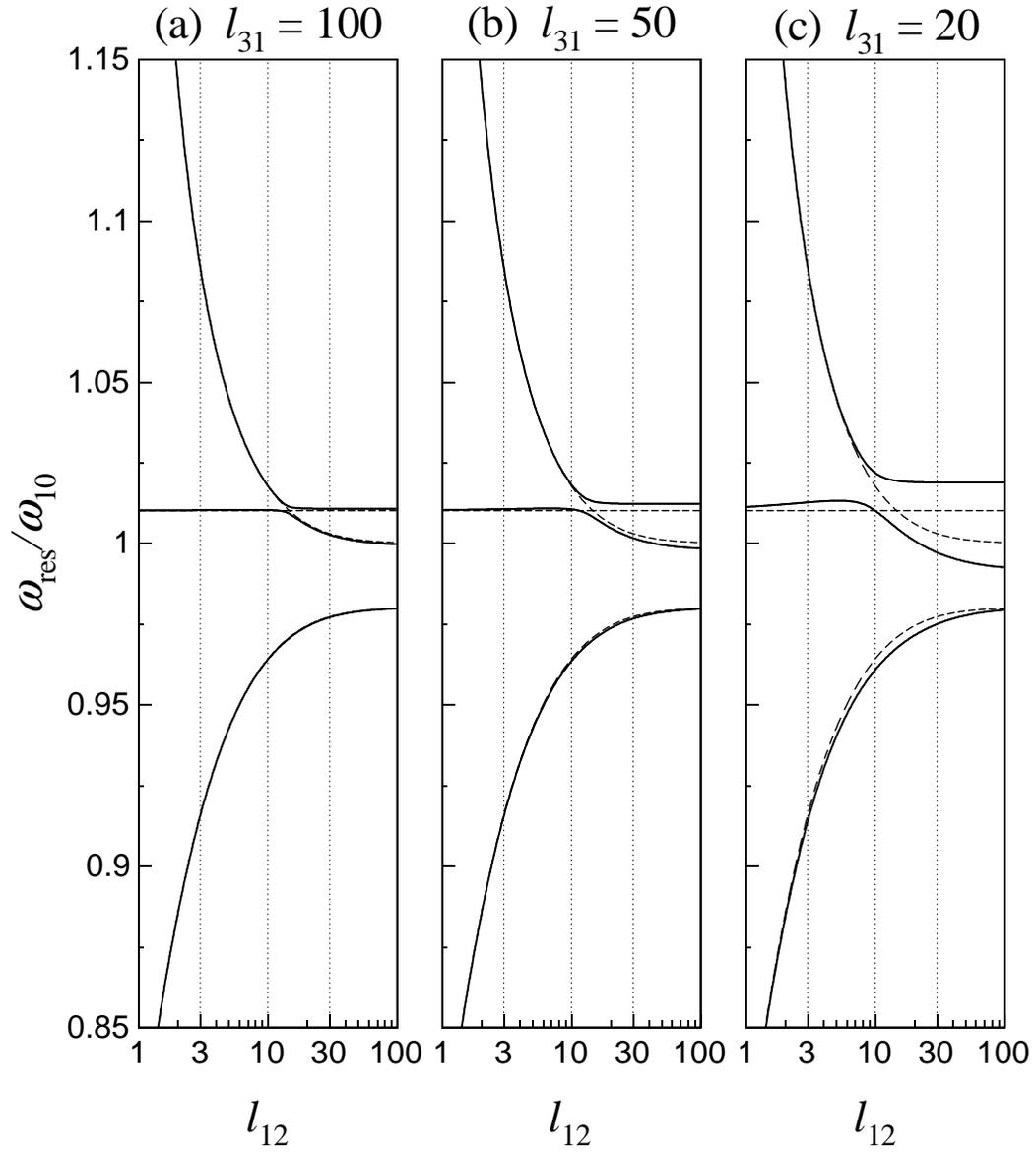}
\caption
{Same as Fig.~\ref{fig2}, but bubble 3 is smaller than the others. The bubbles 
are aligned as shown in Fig.~\ref{fig3}(b).}
\label{fig4}
\end{figure}

An avoided crossing is also observed when bubble 3 is smaller than the 
others. Figure \ref{fig4} shows the resonance frequencies when $R_{30} 
=49.5\;{\rm \mu m}$. Here the bubbles are arranged as illustrated in Fig.~\ref{fig3}(b). If bubble 3 is so large or so small that its monopole resonance 
frequency does not cross with a resonance frequency of the double-bubble 
system, no distinct avoided crossing is observed, though this situation is 
not shown here.

\section{State exchange in the avoided crossing region}
\label{sec4}
To manifest a state exchange like that which the coupled bubbles experience 
through the avoided crossing, we examined the oscillation phases of the 
bubbles. In bubble dynamics, the phase of radial oscillation plays important 
roles in many situations, including acoustic levitation \cite{ref26,ref27,ref28}, bubble-bubble interaction \cite{ref18}, and 
multibubble sonoluminescence 
\cite{ref29}, and hence an accurate understanding of it is crucial. In fact, by 
carefully examining the oscillation phases of two coupled bubbles for weak 
driving, we have recently succeeded in presenting a novel interpretation, 
which may be more accurate than previous ones, of the sign reversal of the 
secondary Bjerknes force \cite{ref18,ref19}, a paradoxical phenomenon that is 
considered to be the cause of the stable structure formation of bubbles in a 
weak acoustic field \cite{ref24,ref30}. In that discussion, it was 
suggested that the transition frequencies seem to be essential components 
for gaining an accurate understanding of the phenomenon, since the sign 
reversal takes place at the transition frequencies that cannot be obtained 
by resonance-frequency analysis. In the present paper, we show by examining 
the oscillation phases that the bubbles exchange their oscillation states 
through the avoided crossing. As shown later, the point at which the state 
exchange occurs can be clearly detected by observing the transition 
frequencies.

\begin{figure}[htbp]
\includegraphics[width=13.5cm]{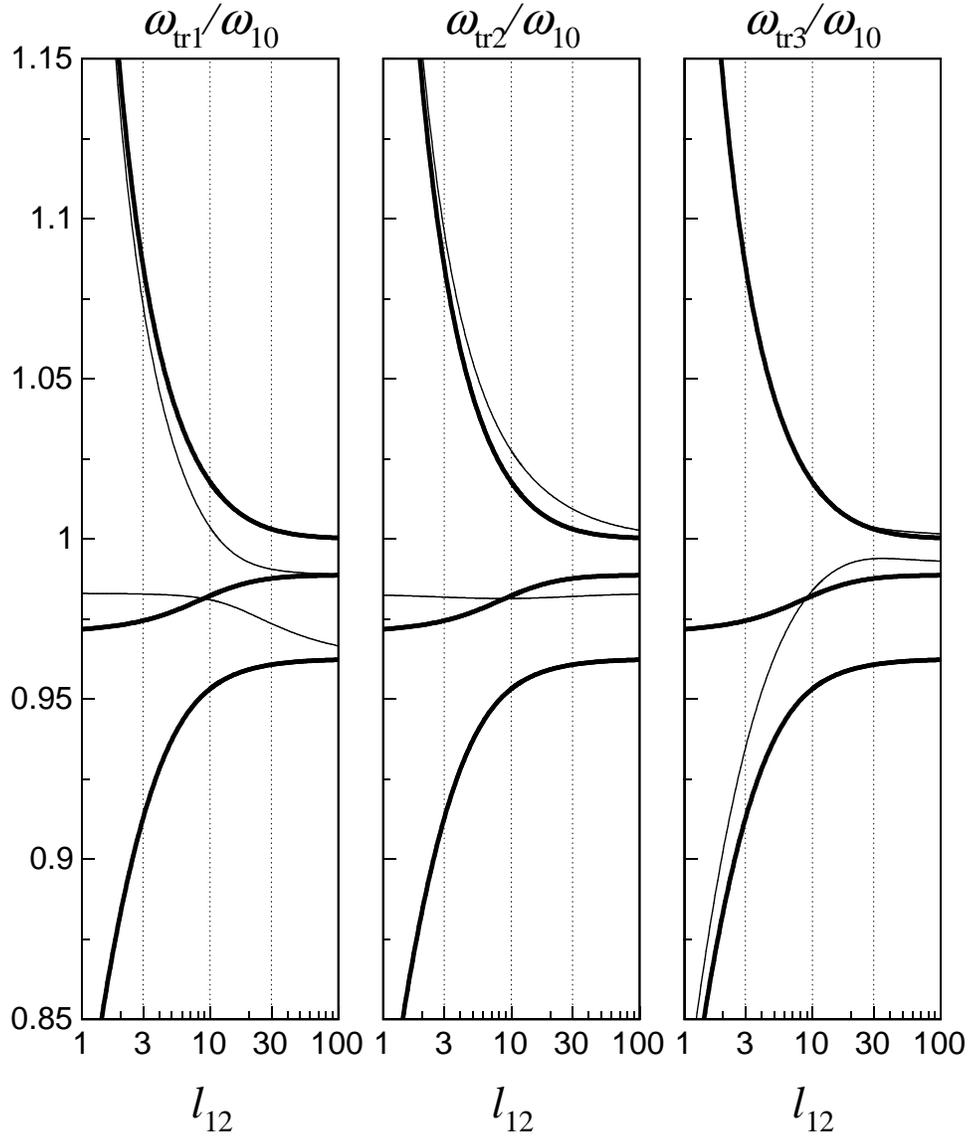}
\caption
{Transition frequencies $\omega _{{\rm tr}} $ (rad/s) of three 
coupled bubbles for $\delta _i \approx 0$ with $l_{23} =20$ normalized by 
$\omega _{10} $ (rad/s), as functions of the normalized separation distance 
$l_{12} $. $\omega _{{\rm tr}\,i} $ denotes the transition frequencies of 
bubble $i$.}
\label{fig5}
\end{figure}

Figure \ref{fig5} shows the transition frequencies for $(R_{10} ,R_{20} ,R_{30} 
)=(50\;{\rm \mu m},51\;{\rm \mu m},51.5\;{\rm \mu m})$ with $l_{23} =20$. 
The thick lines denote the transition frequencies that correspond to the 
resonance frequencies already shown in Fig.~\ref{fig2}(c). As expected from the 
mathematical proof presented in \cite{ref17}, the bubbles have up to five 
transition frequencies, all of which invert the oscillation phase of the 
corresponding bubble. It is worth noting that in each panel of Fig.~\ref{fig5} 
the second-highest resonance frequency (denoted below by $\omega _{2{\rm nd}} 
)$ crosses once with a transition frequency in the avoided crossing region. 
Such crossings have not been found in double-bubble systems \cite{ref16,ref18}. 
In the following discussion, we focus our attention on the phase 
properties of the bubbles in this region to elucidate what happens around 
the intersecting points.

\begin{figure}[htbp]
\includegraphics[width=14cm]{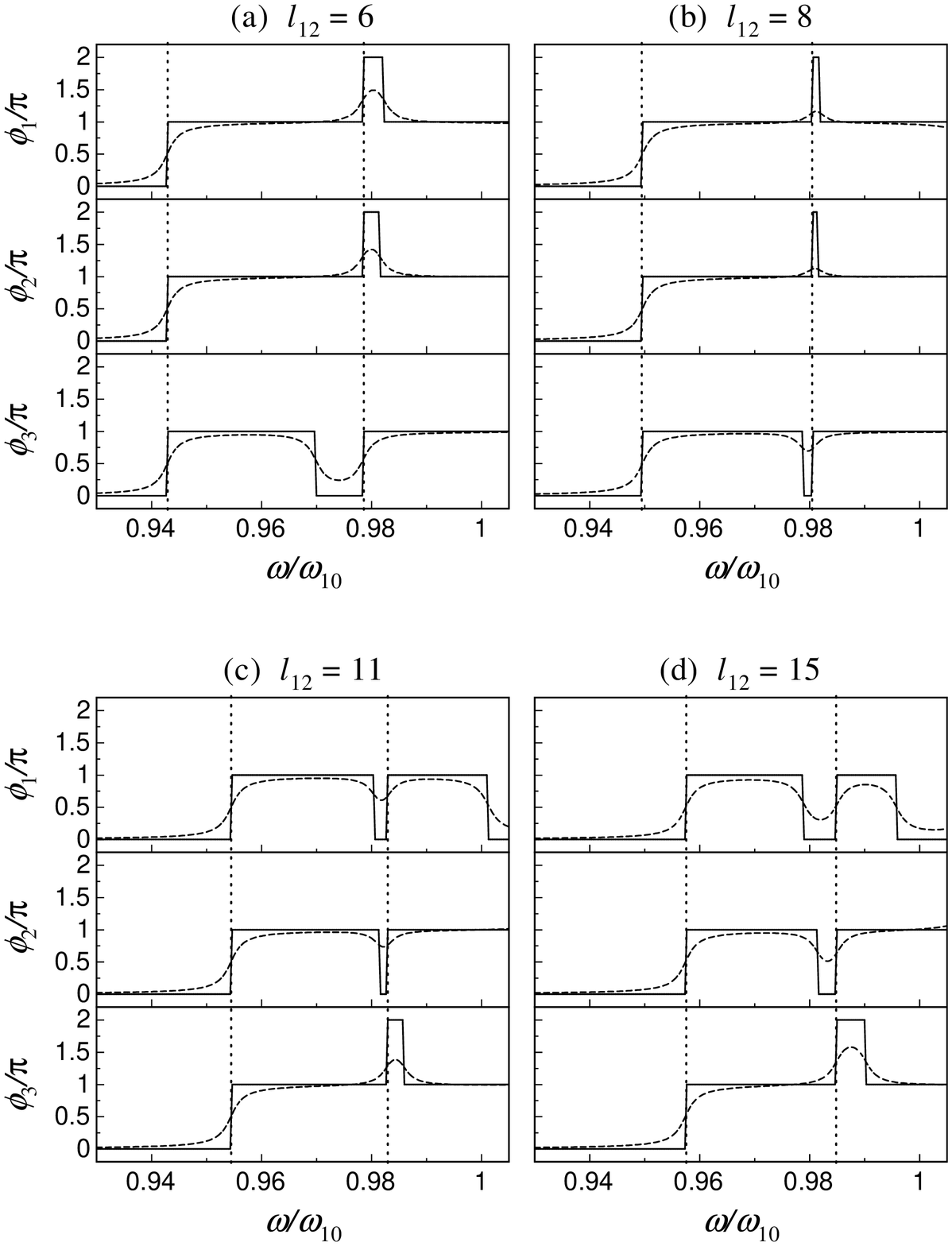}
\caption
{Phase delays $\phi _i $ (rad) normalized by $\pi $ as functions of 
$\omega /\omega _{10} $ for different $l_{12} $ [(a), (b): for $l_{12} 
<l_{\rm int} $, (c), (d): for $l_{12} >l_{\rm int} $]. The solid and the dashed 
curves denote $\phi _i $ for negligible and non-negligible damping, 
respectively, and the vertical dotted lines indicate the two 
lowest resonance frequencies (the higher is thus the second-highest 
resonance frequency $\omega _{2{\rm nd}} )$.}
\label{fig6}
\end{figure}

The phase delays $\phi _i $ for different $l_{12} $ as functions of $\omega 
/\omega _{10} $ are shown by the solid lines in Fig.~\ref{fig6}. In the 
computation of $\phi _i $, we used very small but nonzero $\delta _i $ to obtain continuous results. Figures \ref{fig6}(a, b) and \ref{fig6}(c, d), 
respectively, show 
$\phi _i $ for $l_{12} $ smaller and larger than the intersecting point 
$l_{12} =l_{\rm int} $ ($\approx $8.89). Here, we only displayed $\phi _i $ in 
the frequency range around the two avoided crossing resonance frequencies. 
The vertical dotted lines indicate the two lowest resonance frequencies. As 
in double-bubble cases \cite{ref18,ref19}, at the resonance frequencies the 
phase delays of all bubbles shift simultaneously by $+\pi $, whereas at the 
remaining transition frequencies only one phase delay shifts by $-\pi $.

The $\phi _i $-curves, as can be clearly seen in the figures, have different 
convexities on different sides of the intersecting point. For $l_{12} $ 
smaller than $l_{\rm int} $, at $\omega _{2{\rm nd}} $, $\phi _1 $ and $\phi _2 
$ shift from $\pi $ to $2\pi $ but $\phi _3 $ shifts from $0$ to $\pi $ as 
$\omega $ increases. For $l_{12} $ larger than $l_{\rm int} $, on the other 
hand, an opposite tendency is seen; $\phi _1 $ and $\phi _2 $ shift from $0$ 
to $\pi $ but $\phi _3 $ shifts from $\pi $ to $2\pi $. That is, a kind of 
state exchange takes place between bubble 3 and the other two bubbles at the 
intersecting point.

Regarding the relationship between the state exchange and the phase 
properties, in the frequency range around $\omega _{2{\rm nd}} $, bubble 3 
oscillates out-of-phase with the other bubbles regardless of whether $l_{12} 
<l_{\rm int} $ or $l_{12} >l_{\rm int} $, although the individual phase delays 
experience rapid shifts at $\omega _{2{\rm nd}} $. This means that the 
state exchange cannot be perceived accurately by observing whether the 
bubbles oscillate in-phase or out-of-phase with each other or by observing 
the sign of the secondary Bjerknes force, which is determined by the cosine 
of the phase difference between two bubbles \cite{ref31,ref32}. Just the 
individual phase delays (or transition frequencies) should be examined.

In the $\phi _i $-curves, we can find several similarities with 
double-bubble cases. Bubbles 1 and 2, or bubble 3, have a phase delay 
greater than $\pi $ in the frequency range from $\omega _{2{\rm nd}} $ to a 
certain higher frequency (equal to the next-higher transition frequency of 
the corresponding bubble). A similar observation can be found for 
double-bubble systems \cite{ref18,ref19}. In Ref.~\cite{ref18} we discovered 
and elucidated that such a large phase delay can appear when two bubbles 
interact with each other through sound. In the double-bubble case, the 
larger one of the two bubbles has a phase delay greater than $\pi $ in the 
frequency range between the higher of two resonance frequencies and the 
highest of the transition frequencies of the bubble. We can, for a wider 
frequency range, also find a similarity between the double- and 
triple-bubble cases. In the frequency range $\omega /\omega _{10} < 0.995$, 
the profiles of $\phi _1 $ and $\phi _2 $ for 
$l_{12} <l_{\rm int} $ and that of $\phi _3 $ for $l_{12} >l_{\rm int} $ are very 
similar to the profile of the phase delay of the larger bubble in a 
double-bubble system; those phase delays first exhibit two sharp rises and 
then one sharp fall as $\omega $ increases. Also, the profiles of the 
remaining phase delays are very similar to that of the phase delay of the 
smaller bubble in a double-bubble system, exhibiting one sharp rise, one 
sharp fall, and then one sharp rise. This seems to indicate that in the 
frequency range considered, for $l_{12} <l_{\rm int} $ bubbles 1 and 2 act as 
``larger bubbles'' while bubble 3 acts as a ``smaller bubble,'' but for 
$l_{12} >l_{\rm int} $ each bubble acts in the opposite way; that is, the 
physical roles that the bubbles play are exchanged through the avoided 
crossing. This observation could also be interpreted as a result of the 
change of a physical meaning of $\omega _{2{\rm nd}} $. As illustrated in 
Fig.~\ref{fig2}, $\omega _{2{\rm nd}} $ is a hybrid of two resonance frequencies 
having different origins. We assume here that the origin, or the principal 
origin, of each avoided crossing resonance frequency is switched at $l_{\rm int} 
$. This assumption allows us to consider that $\omega _{2{\rm nd}} $ for 
$l_{12} <l_{\rm int} $, for example, is the resonance frequency whose principal 
origin is bubble 3. This suggestion is consistent, not only with the 
observation for large $l_{23} $ where the origin of each resonance frequency 
is relatively clear, but also with the above speculation that bubbles 3 acts 
as a ``smaller bubble'' for $l_{12} <l_{\rm int} $, because $\omega _{2{\rm nd}}$ 
is higher than the lowest resonance frequency that is one of 
the two avoided crossing resonance frequencies. The observation for $l_{12} 
>l_{\rm int} $ can be interpreted in a similar manner.

Lastly, we briefly examine how the damping affects the state exchange. For 
the damping coefficient, we use the value for viscous damping,
\begin{equation}
\label{eq13}
\delta _i =\frac{4\mu }{\rho R_{i0}^2 }
\end{equation}
with viscosity $\mu =1.002\times 10^{-3}$ kg/(m s). The dashed curves in 
Fig.~\ref{fig6} show the phase delays in the damped case. The viscous effect smoothes 
the phase profiles, but the convexity of the curves is not altered from that 
for $\delta _i \approx 0$, as in the double-bubble cases \cite{ref18,ref19}. 
The state exchange is clearly detected even in the present case. The 
qualitative tendencies of the phase delays are not changed by the viscous 
damping.

\section{Conclusion}
\label{sec5}
We have shown theoretically that avoided crossings can be observed in the 
resonance frequencies of acoustically coupled gas bubbles plotted as 
functions of the separation distances. A state exchange taking place between 
the bubbles in the avoided crossing region has been clearly exhibited by 
examining the oscillation phases and transition frequencies of the coupled 
bubbles. We have clarified that the state exchange is perceived by observing 
the individual oscillation phases of the bubbles, not by observing whether 
the bubbles oscillate in-phase or out-of-phase with each other. Since the 
individual phase (or more properly, the phase difference between a bubble 
and the external sound) determines the sign of the primarily Bjerknes force 
\cite{ref26,ref27,ref28} acting on the corresponding bubble, this state 
exchange should play a role in, e.g., acoustic levitation using the force. The 
results of this study suggest that the transition frequencies introduced in 
Ref.~\cite{ref16} 
can be a useful tool for detecting the state exchange, which takes 
place at the separation distance where an avoided crossing resonance 
frequency crosses with a transition frequency that is not a resonance 
frequency. Though we only considered triple-bubble systems in a linear 
arrangement, extensions to systems containing a larger number of bubbles and 
in different arrangements may be straightforward. 
Also, nonlinear effects on the avoided crossings and oscillation phases could 
be examined using nonlinear models \cite{ref20,ref21,ref29,ref33}. 
As with other physical 
systems, the avoided crossings in acoustically coupled bubbles might be 
real.

\begin{acknowledgments}
The author thanks Dr.~Akemi Nishida for valuable comments. This work was 
supported by a Grant-in-Aid for Young Scientists (B) (17760151) from the 
Ministry of Education, Culture, Sports, Science, and Technology of Japan.
\end{acknowledgments}

\appendix

\section{}
\label{secA}
High-order nonlinear models for $N$ pulsating bubbles in a liquid have been 
proposed, in which terms proportional to $D_{i\,j}^{-k} $ ($k\ge 2)$ appear 
that involve the translational velocities of the bubbles \cite{ref21,ref33}. In 
Ref.~\cite{ref33}, for example, Doinikov derived the following 
model equation for $N$ spherical bubbles:
\begin{equation}
\label{eq14}
R_i \ddot {R}_i +\frac{3}{2}\dot {R}_i^2 -\frac{P_i }{\rho }=\frac{{\rm {\bf 
\dot {p}}}_i^2 }{4}-\sum\limits_{j=1, j\ne i}^N {\left\{ {\frac{R_j^2 \ddot 
{R}_j +2R_j \dot {R}_j^2 }{D_{i\,j} }+H_{i\,j} } \right\}} ,
\end{equation}
\begin{eqnarray}
\label{eq15}
\frac{1}{3}R_i {\rm {\bf \ddot {p}}}_i +\dot {R}_i {\rm {\bf \dot {p}}}_i 
=\frac{{\rm {\bf F}}_i }{2\pi \rho R_i^2 } &+& \sum\limits_{j=1, j\ne i}^N 
\left\{ -\frac{1}{D_{i\,j}^2 }(R_i R_j^2 \ddot {R}_j +2R_i R_j \dot {R}_j^2 
+\dot {R}_i \dot {R}_j R_j^2 ){\rm {\bf t}}_{i\,j} \right. \nonumber \\ 
&& -\frac{R_j^2 }{2D_{i\,j}^3 }[R_i R_j {\rm {\bf \ddot {p}}}_j +(\dot {R}_i 
R_j +5R_i \dot {R}_j ){\rm {\bf \dot {p}}}_j ] \nonumber \\ 
&& + \left. \frac{3R_j^2 }{2D_{i\,j}^3 }\{{\rm {\bf t}}_{i\,j} \cdot [R_i R_j {\rm {\bf \ddot {p}}}_j +(\dot {R}_i R_j +5R_i \dot {R}_j ){\rm {\bf \dot {p}}}_j ]\}{\rm {\bf t}}_{i\,j} \right\} ,
\end{eqnarray}
with
\begin{eqnarray}
\label{eq16}
H_{i\,j} &\equiv& -\frac{R_j^2 }{2D_{i\,j}^2 }(R_j {\rm {\bf \ddot {p}}}_j 
+\dot {R}_j {\rm {\bf \dot {p}}}_i +5\dot {R}_j {\rm {\bf \dot {p}}}_j 
)\cdot {\rm {\bf t}}_{i\,j} \nonumber \\ 
&&-\frac{R_j^3 }{4D_{i\,j}^3 }\left[ {{\rm {\bf \dot {p}}}_j \cdot ({\rm {\bf 
\dot {p}}}_i +2{\rm {\bf \dot {p}}}_j )-3({\rm {\bf \dot {p}}}_j \cdot {\rm 
{\bf t}}_{i\,j} )[{\rm {\bf t}}_{i\,j} \cdot ({\rm {\bf \dot {p}}}_i +2{\rm 
{\bf \dot {p}}}_j )]} \right],
\end{eqnarray}
\[
{\rm {\bf t}}_{i\,j} \equiv \frac{{\rm {\bf p}}_j -{\rm {\bf p}}_i}{D_{i\,j} },
\]
\begin{equation}
\label{eq17}
P_i \equiv \left( {P_0 +\frac{2\sigma }{R_{i0} }} \right)\left( 
{\frac{R_{i0} }{R_i }} \right)^{3\gamma }-\frac{2\sigma }{R_i }-\frac{4\mu 
\dot {R}_i }{R_i }-P_0 -p_{{\rm ex}} ,
\end{equation}
where $R_i $ and ${\rm {\bf p}}_i $ are the instantaneous radius and 
position vector, respectively, of bubble $i$, ${\rm {\bf F}}_i $ denotes 
external forces on bubble $i$, ${\rm {\bf t}}_{i\,j} $ is a unit vector, 
$\gamma$ is the specific heat ratio of the gas inside the bubbles, and 
$\mu $ is the viscosity. Here we showed only the incompressible version, 
though Doinikov also derived a model for bubbles in a compressible liquid. 
Equations (\ref{eq14}) and (\ref{eq15}) represent the volume oscillation of 
bubble $i$ and its translational motion, respectively. The linear coupled 
oscillator model used in the present study is recovered from Eq.~(\ref{eq14}) 
by truncating the high-order terms $H_{i\,j} $ and assuming weak driving and 
$\gamma = \kappa _i$.

Since the velocity field forming around a pulsating sphere is proportional 
to $1/r^2$, where $r$ is the distance measured from the center of the 
sphere, the truncated terms $H_{i\,j} $, which are composed of the 
translational velocities ${\rm {\bf \dot {p}}}_i $, might be considered to 
be of fourth, or higher, order with respect to the inverse of the separation 
distances. This speculation is consistent with the suggestion by Harkin et 
al.~for double-bubble systems \cite{ref20}.

Equation (\ref{eq14}) further suggests that under the assumption of ${\rm {\bf 
\dot {p}}}_i \approx {\rm {\bf 0}}$ one cannot construct a linear model that 
has higher-order accuracy than that of Eq.~(\ref{eq1}), since this assumption 
makes the high-order terms inaccurate.

\section{}
\label{secB}
For the convenience of readers, we show the concrete forms of $\left| {\rm 
{\bf A}} \right|$ and ${\rm {\bf C}}\,{\rm {\bf I}}$:
\begin{eqnarray}
\label{eq18}
\frac{\vert {\rm {\bf A}}\vert }{R_{10} R_{20} R_{30} } &=& L_1 L_2 L_3 
+s_{21} s_{32} s_{13} +s_{12} s_{23} s_{31} \nonumber \\ 
&& -L_1 (M_2 M_3 +s_{23} s_{32} )-L_2 (M_3 M_1 +s_{31} s_{13} )-L_3 
(M_1 M_2 +s_{12} s_{21} ) \nonumber \\ 
&& +{\rm i} \left[ M_1 M_2 M_3 -M_1 (L_2 L_3 -s_{23} s_{32} )-M_2 (L_3 
L_1 -s_{31} s_{13} ) \right. \nonumber \\ 
&& - \left. M_3 (L_1 L_2 -s_{12} s_{21} ) \right],
\end{eqnarray}
\[
{\rm {\bf C}}\,{\rm {\bf I}}=(c_1 ,c_2 ,c_3 )^T,
\]
\begin{eqnarray}
\label{eq19}
\frac{c_i }{R_{j0} R_{k0} } &=& (L_j -s_{i\,j} )(L_k -s_{i\,k} )+(s_{i\,j} 
-s_{k\,j} )(s_{j\,k} -s_{i\,k} )-M_j M_k \nonumber \\ 
&& +{\rm i}\left[ {M_j (s_{i\,k} -L_k )+M_k (s_{i\,j} -L_j )} \right] \nonumber \\ 
&& \mbox{for }(i,j,k) = (1,2,3),(2,3,1),\mbox{ or }(3,1,2),
\end{eqnarray}
where
\[
L_i \equiv X-\omega _{i0}^2 ,
\]
\[
M_i \equiv \omega \delta _i ,
\]
\[
s_{i\,j} \equiv \frac{R_{j0} }{D_{i\,j} }X.
\]
For $\delta _i \approx 0$, Eqs.~(\ref{eq18}) and (\ref{eq19}) reduce, respectively, to
\begin{eqnarray}
\label{eq20}
\frac{\vert {\rm {\bf A}}\vert }{R_{10} R_{20} R_{30} } &\approx& L_1 L_2 L_3 +s_{21} s_{32} s_{13} +s_{12} s_{23} s_{31} \nonumber \\ 
&& -L_1 s_{23} s_{32} -L_2 s_{31} s_{13} -L_3 s_{12} s_{21} ,
\end{eqnarray}
\begin{equation}
\label{eq21}
\frac{c_i }{R_{j0} R_{k0} }\approx (L_j -s_{i\,j} )(L_k -s_{i\,k} )+(s_{i\,j} -s_{k\,j} )(s_{j\,k} -s_{i\,k} ).
\end{equation}
Equation (\ref{eq20}) and the real part of Eq.~(\ref{eq18}) are cubic functions and Eq.~(\ref{eq21}) 
and the real part of Eq.~(\ref{eq19}) are quadratic functions in terms of $X$. The 
imaginary parts of Eqs.~(\ref{eq18}) and (\ref{eq19}) can be written in a form of $\omega 
f(X)$, where $f$ is quadratic in Eq.~(\ref{eq18}) and linear in 
Eq.~(\ref{eq19}). (As proved theoretically in Ref.~\cite{ref17}, the imaginary 
parts are composed of terms of odd orders with respect to $M$ that are 
proportional to $\omega X^n$ with 
$n$ being a positive integer.)

\newpage 

\newpage 

\newpage 

\newpage 

\newpage 

\newpage 


\begin{thebibliography}{}
\bibitem{ref1}
In the fields of vibration engineering and some others, avoided 
crossings are called ``curve veering,'' ``eigenvalue veering,'' or 
``(natural) frequency loci veering.'' They are also sometimes called 
``avoided level crossings.''

\bibitem{ref2}
J. R. Walkup, M. Dunn, D. K. Watson, and T. C. Germann, Phys. Rev. A {\bf 58}, 4668 (1998).

\bibitem{ref3}
T. Timberlake and L. E. Reichl, Phys. Rev. A {\bf 59}, 2886 (1999).

\bibitem{ref4}
S. Li and E. J. Heller, Phys. Rev. A {\bf 67}, 032712 (2003).

\bibitem{ref5}
S. D\"{u}rr, T. Volz, A. Marte, and G. Rempe, Phys. Rev. Lett. {\bf 92}, 
020406 (2004).

\bibitem{ref6}
Y. Osaki, Publ. Astron. Soc. Jpn. {\bf 27}, 237 (1975).

\bibitem{ref7}
D. Gondek and J. L. Zdunik, Astron. Astrophys. {\bf 344}, 117 (1999).

\bibitem{ref8}
S. Shaik, A. Ioffe, A. C. Reddy, and A. Pross, J. Am. Chem. Soc. {\bf 116}, 262 (1994).

\bibitem{ref9}
C. Zhu and H. Nakamura, Chem. Phys. Lett. {\bf 274}, 205 (1997).

\bibitem{ref10}
J. R. Kuttler and V. G. Sigillito, J. Sound Vib. {\bf 75}, 585 (1981).

\bibitem{ref11}
N. C. Perkins and C. D. Mote, Jr., J. Sound Vib. {\bf 106}, 451 (1986).

\bibitem{ref12}
C. Pierre, J. Sound Vib. {\bf 126}, 485 (1988).

\bibitem{ref13}
H. H. Yoo and S. H. Shin, J. Sound Vib. {\bf 212}, 807 (1998).

\bibitem{ref14}
R. S. Langley, Proc. R. Soc. Lond. A {\bf 455}, 3325 (1999).

\bibitem{ref15}
A. A. Mailybaev, O. N. Kirillov, and A. P. Seyranian, J. Phys. A {\bf 38}, 1723 (2005).

\bibitem{ref16}
M. Ida, Phys. Lett. A {\bf 297}, 210 (2002).

\bibitem{ref17}
M. Ida, J. Phys. Soc. Jpn. {\bf 71}, 1214 (2002).

\bibitem{ref18}
M. Ida, Phys. Rev. E {\bf 67}, 056617 (2003).

\bibitem{ref19}
M. Ida, J. Phys. Soc. Jpn. {\bf 73}, 3026 (2004).

\bibitem{ref20}
A. Harkin, T. J. Kaper, and A. Nadim, J. Fluid Mech. {\bf 445}, 377 
(2001).

\bibitem{ref21}
H. Takahira, T. Akamatsu, and S. Fujikawa, JSME Int. J. Ser. B {\bf 37}, 297 (1994).

\bibitem{ref22}
C. Feuillade, J. Acoust. Soc. Am. {\bf 98}, 1178 (1995).

\bibitem{ref23}
A. Shima, Trans. ASME, J. Basic Eng. {\bf 93}, 426 (1971).

\bibitem{ref24}
E. A. Zabolotskaya, Sov. Phys. Acoust. {\bf 30}, 365 (1984).

\bibitem{ref25}
For such a large separation distance, in practice the time delay effect 
due to the finite sound speed of the surrounding liquid should be taken into 
account, though we only present an idealized result for the sake of 
simplicity.

\bibitem{ref26}
F. G. Blake, Jr., J. Acoust. Soc. Am. {\bf 21}, 551 (1949).

\bibitem{ref27}
T. J. Matula, Phil. Trans. R. Soc. Lond. A {\bf 357}, 225 (1999).

\bibitem{ref28}
I. Akhatov, R. Mettin, C. D. Ohl, U. Parlitz, and W. Lauterborn, Phys. Rev. E {\bf 55}, 3747 (1997).

\bibitem{ref29}
R. Mettin, I. Akhatov, U. Parlitz, C. D. Ohl, and W. Lauterborn, Phys. Rev. E {\bf 56}, 2924 (1997).

\bibitem{ref30}
A. A. Doinikov and S. T. Zavtrak, J. Acoust. Soc. Am. {\bf 99}, 3849 
(1996).

\bibitem{ref31}
W. Lauterborn, T. Kurz, R. Mettin, and C. D. Ohl, Adv. Chem. Phys. {\bf 
110}, 295 (1999).

\bibitem{ref32}
L. A. Crum, J. Acoust. Soc. Am. {\bf 57}, 1363 (1975).

\bibitem{ref33}
A. A. Doinikov, J. Acoust. Soc. Am. {\bf 116}, 821 (2004).
\end{thebibliography}
\end{document}